\documentclass[12pt,letter]{article}
\usepackage{graphicx, epsfig, color}
\usepackage{amsmath,amssymb,hyperref}
\usepackage{subfigure}
\def\eslt{\not\!\!{E_T}}
\def\to{\rightarrow}

\def\bi{\begin{itemize}}
\def\ei{\end{itemize}}

\def\tchi{\tilde\chi}

\def\sps1ap{SPS1a$^\prime$}
\def\c1p{C1$^\prime$}

\def\tst{\tilde t}

\def\tg{\tilde g}

\def\alt{\lesssim}
\def\agt{\gtrsim}
\def\be{\begin{equation}}  
\def\ee{\end{equation}}  
\def\bea{\begin{eqnarray}}  
\def\eea{\end{eqnarray}}  
\def\beas{\begin{eqnarray*}}  
\def\eeas{\end{eqnarray*}}

\begin{document}
\begin{titlepage}
\begin{flushright}
OUHEP-240731
\end{flushright}

\vspace{0.5cm}
\begin{center}
  {\Large \bf Minding the gap: testing natural anomaly-mediated\\
    SUSY breaking at high luminosity LHC
}\\ 
\vspace{1.0cm} \renewcommand{\thefootnote}{\fnsymbol{footnote}}
{\large Howard Baer$^1$\footnote[1]{Email: baer@nhn.ou.edu }, 
Vernon Barger$^2$\footnote[2]{Email: barger@pheno.wisc.edu },
Jessica Bolich$^1$\footnote[3]{Email: Jessica.R.Bolich-1@ou.edu}\\
Juhi Dutta$^1$\footnote[4]{Email: juhi.dutta@ou.edu} and
Dibyashree Sengupta$^3$\footnote[5]{Email: Dibyashree.Sengupta@lnf.infn.it}
}\\ 
\vspace{1.0cm} \renewcommand{\thefootnote}{\arabic{footnote}}
{\it 
$^1$Department of Physics and Astronomy,
University of Oklahoma, Norman, OK 73019, USA \\
}
{\it 
$^2$Department of Physics,
University of Wisconsin, Madison, WI 53706, USA \\
}
{\it
  $^3$ INFN, Laboratori Nazionali di Frascati,
Via E. Fermi 54, 00044 Frascati (RM), Italy}
\end{center}

\vspace{0.4cm}
\begin{abstract}
\noindent 
While the minimal anomaly-mediated SUSY breaking model (mAMSB) seems ruled
out by constraints on Higgs mass, naturalness and wino dark matter,
a slightly generalized version dubbed natural AMSB (nAMSB) remains both viable
and compelling.
Like mAMSB, nAMSB features winos as the lightest gauginos, but unlike mAMSB,
nAMSB allows a small $\mu$ parameter so that higgsinos are the lightest
of electroweakinos (EWinos).
nAMSB spectra depend on the input value of gravitino mass $m_{3/2}$,
where the lower range of $m_{3/2}$ is excluded by LHC gluino pair searches
while a higher $m_{3/2}$ band is excluded by LHC limits on wino pair
production followed by boosted hadronic wino decays.
A remaining intermediate gap in $m_{3/2}$ values
remains allowed by present LHC searches, but appears to be completely
explorable by high luminosity ugrades of LHC (HL-LHC).
We explore a variety of compelling discovery channels that may allow one
to close the intermediate gap in $m_{3/2}$ values:
1. same-sign diboson$+\eslt$ (SSdB) production arising from wino pair
production, leading to same-sign dileptons plus missing $E_T$,
2. trilepton production arising from wino pair production and
3. soft dilepton plus jet events from higgsino pair production,
4. top-squark pair production.
From our signal-to-background analysis along a nAMSB model line,
we expect HL-LHC to either discover or rule out
the nAMSB model with 3000 fb$^{-1}$ of integrated luminosity.
\end{abstract}

\end{titlepage}

\section{Introduction}
\label{sec:intro}

In gravity-mediated supersymmetry breaking models\cite{Nilles:1983ge,Chung:2003fi},
one posits a visible sector (usually consisting of Minimal Supersymmetric
Standard Model or MSSM fields) and a hidden sector which allows for
supersymmetry breaking to occur.
Under supergravity breaking\cite{Cremmer:1982en}, some hidden sector
field $X$ develops a SUSY breaking vev $F_X$ so that the gravitino acquires a
mass $m_{3/2}\sim F_X/m_P$ which can be of order the weak
scale $m_{weak}$ for $F_X\sim m_{hidden}^2\sim (10^{11}\ {\rm GeV})^2$.
This can then lead to the generation of visible sector
soft SUSY breaking scalar mass-squared $m_{\phi}^2$, gaugino masses $m_\lambda$,
trilinear $(A)$ and bilinear $B$ terms all of order $m_{3/2}$,
which can lead to a 't Hooft technically natural solution\cite{tHooft:1979rat} of the
Standard Model (SM) big hierarchy problem (BHP)\cite{Witten:1981nf}
and a practical natural solution to the MSSM little hierarchy problem
(LHP)\cite{Baer:2023cvi}.
But how can the exponentially suppressed scale $\sqrt{F_X}\ll m_P$ arise?
An attractive framework lies in dynamical SUSY breaking (DSB)\cite{Dine:2010cv}
where non-perturbative breaking, say via hidden sector gaugino
condensation\cite{Nilles:1990zd} where a gauge group (say $SU(N)$)
becomes strongly interacting at the intermediate scale leading to
$m_{hidden}\sim m_P\exp (-8\pi^2/g_h^2)$, and the hidden sector mass scale
arises via dimensional transmutation.

A conundrum with this approach is that gaugino masses and $A$-terms
require hidden sector singlets to develop vevs, and singlets do not seem to
have a place within the hidden sector gauge theory\cite{Affleck:1984xz}.
Nonetheless, it was pointed out by Randall-Sundrum\cite{Randall:1998uk} and Giudice {\it et al.}\cite{Giudice:1998xp} that these terms
could arise with loop suppressed values which trace back to the superconformal
anomaly, thus dubbed as anomaly-mediated SUSY breaking (AMSB) terms.
In this case, gaugino masses arise as
\be
M_i =\frac{\beta_{g_i}}{g_i}m_{3/2}
\ee
where $i$ labels the corresponding gauge group and and $\beta_{g_i}$ is
its $\beta$ function. $A_f$ terms develop as $= \frac{\beta_f}{f}m_{3/2}$
where $\beta_f$ is the beta function for the corresponding
superpotential Yukawa coupling. Anomaly-mediated contributions to scalar
masses can also be derived and can be found in textbooks such as
Ref. \cite{Baer:2006rs}.

In cases where tree-level gravity-mediated soft terms are suppressed\footnote{Originally, sequestering the hidden sector from the visible sector via brane separation in extra dimensions was suggested\cite{Randall:1998uk}.
There is still on-going debate
regarding viable sequestration methods in SUSY models\cite{Dine:2007me}.},
then the AMSB contributions to soft SUSY breaking terms may dominate,
leading to models with distinct phenomenological predictions.
Famously, since gaugino masses are proportional to their gauge group beta
functions, then one expects the $SU(3)_C$, $SU(2)_L$ and $U(1)_Y$ gaugino masses
at the weak scale to occur in the ratios $M_3:M_2:M_1\sim 8:1:3$
so that the wino is expected to be the lightest gaugino
(as opposed to models with gaugino mass unification,
where the bino is expected as the lightest gaugino).
The advent of AMSB led to formulation of a minimal AMSB model
(mAMSB)\cite{Gherghetta:1999sw,Feng:1999hg} defined by parameters
\be
m_0,\ m_{3/2},\ \tan\beta,\ sign (\mu )\ \ \ (mAMSB)
\ee
where a so-called {\it bulk} scalar mass contribution $m_0$ was postulated
to avoid tachyonic sleptons. The magnitude of the superpotential $\mu$
parameter was tuned so as to generate the observed value of $m_Z=91.2$ GeV.
The mAMSB model featured a wino-like lightest SUSY particle (LSP)
and hence wino-like WIMP dark matter. The thermally underproduced
wino-like WIMP abundance could be augmented to the  measured DM
abundance via moduli field decays in the early universe\cite{Moroi:1999zb}.
The mAMSB model provided a template for many investigations of
how AMSB SUSY would manifest itself phenomenologically\cite{Paige:1999ui,Baer:2000bs,Barr:2002ex}.

Nowadays, the mAMSB model seems highly implausible-- perhaps even excluded--
based on three counts:
\begin{enumerate}
\item The small $A_0\sim 0$ parameter means little mixing of top-squarks
  so that the relatively large measured value of $m_h\simeq 125$ GeV can only be obtained by large, unnatural top-squark soft terms lying in the
  tens-to-hundreds of TeV regime\cite{Arbey:2011ab,Baer:2012mv}.
  Such large stop masses lead to a naturalness problem\cite{Baer:2014ica}.
\item A necessary condition for naturalness in SUSY models is that
  $\mu\sim m_{weak}\simeq m_{W,Z,h}\sim 100$ GeV.
  The large value of $\mu$ which is generated in mAMSB makes the model
  unnatural.
\item Limits on wino-like WIMP dark matter from indirect detection
  experiments effectively exclude wino-like
  LSPs\cite{Cohen:2013ama,Fan:2013faa,Baer:2016ucr} (although such constraints
  can be avoided by postulating mixed axion/wino dark matter
  (two DM particles)\cite{Bae:2015rra}).
\end{enumerate}

In our discussion, it is essential to define what we mean by naturalness.
Here, we adopt the (most conservative) electroweak naturalness measure
$\Delta_{EW}$. This measure depends on the computed value of
$m_Z$ from the scalar potential minimization conditions:
\be m_Z^2/2=\frac{m_{H_d}^2+\Sigma_d^d-(m_{H_u}^2+\Sigma_u^u )
  \tan^2\beta}{\tan^2\beta -1}-\mu^2\simeq -m_{H_u}^2-\mu^2-\Sigma_u^u (\tst_{1,2})
\label{eq:mzs}
\ee 
where the $\Sigma_{u,d}^{u,d}$ terms contain an assortment of loop
corrections which are given in the Appendices of Ref's \cite{Baer:2012cf}
and \cite{Baer:2021tta}. $\Delta_{EW}$ compares the largest term on the
right-hand-side (RHS) of Eq. \ref{eq:mzs} to $m_Z^2/2$.
If one (or more) term(s) on the RHS are far bigger than $m_Z^2/2$, then
some other unrelated term will have to be opposite-sign but nearly same
magnitude to maintain $m_Z=91.2$ GeV.
This is where the finetuning actually occurs in spectra generating
computer codes. Older measures are problematic as explained in
Ref. \cite{Baer:2013gva}. A value $\Delta_{EW}\alt 30$ corresponds to
$\mu\alt 350$ GeV which coincides well with the Agrawal {\it et al.}\cite{Agrawal:1997gf}
window of values (ABDS window) giving rise to anthropic selection of
the magnitude of the weak scale on the string landscape\cite{Baer:2017uvn}.

A new phenomenological model dubbed {\it natural AMSB} (nAMSB)
was proposed in Ref. \cite{Baer:2018hwa} which
incorporated two minor adjustments to mAMSB\footnote{These minor adjustments
  were already suggested in the original RS paper\cite{Randall:1998uk}.}.
First, there is no need for universality amongst the matter scalar masses,
and in fact one of the hallmarks of SUGRA is that such terms are expected to be
{\it non-universal}\cite{Soni:1983rm}.
Thus, independent Higgs bulk terms were postulated
such that $m_{H_u}^2\ne m_{H_d}^2\ne m_0(i)^2$ where $m_0(i)$ denotes
the soft mass for each generation $i$ of matter scalars. The independent
high scale Higgs bulk soft terms can then be traded for the more convenient
weak scale parameters $\mu$ and $m_A$. This allows for smaller values
$\mu\sim 100-350$ GeV as required by weak scale naturalness\cite{Baer:2012up}.
Second, a bulk $A_0$ term is also allowed. Such a bulk $A_0$ term allows
for large stop mixing which then allows for $m_h\sim 125$ GeV but with
top-squark masses in the few TeV range in accord with low $\Delta_{EW}\alt 30$.
The parameter space of the retrofitted nAMSB model is thus given by
\be
m_0(i),\ m_{3/2},\ A_0,\ \tan\beta,\ \mu\ \ {\rm and}\ m_A\ \ \ (nAMSB) .
\ee

The nAMSB model thus allows for $m_h\sim 125$ GeV while $\Delta_{EW}\alt 30$,
all while being in accord with LHC sparticle search constraints.
Like the mAMSB model, the wino is still the lightest gaugino. Unlike mAMSB,
the lightest higgsino is now the lightest electroweakino (EWino),
and in fact the LSP.
The higgsino-like LSP is also thermally underproduced.
However, if naturalness is also required in the QCD sector,
then the PQ solution\cite{Peccei:1977hh,Weinberg:1977ma,Wilczek:1977pj}
to the strong CP problem seems required and we expect mixed
axion/higgsino-like WIMP dark matter\cite{Bae:2013bva}
which avoids indirect dark matter detection constraints.
Given the rearranged EWino mass hierarchy
\be
m(higgsino)\ll m(wino)\ll m(bino)\ll m(gluino)
\ee
 from nAMSB,
the expected collider phenomenology also changes markedly from what was
found in mAMSB. In particular, whereas the lightest neutral
wino was expected to comprise DM in mAMSB, now in nAMSB both the
neutral and charged winos are unstable and decay to vector or Higgs bosons
plus the lighter higgsino states.

In Ref. \cite{Baer:2023ech}, the expected collider phenomenology of nAMSB
was investigated. LHC searches for gluino pair production in the
context of simplified models\cite{CMS:2019zmd,ATLAS:2019vcq} are
expected to apply to nAMSB.
The Run 2 limit that $m_{\tg}\agt 2.2$ TeV implies that
$m_{3/2}\agt 90$ TeV, thus providing a lower limit on expected
$m_{3/2}$ values. Also, recent ATLAS searches for boosted
hadronically-decaying EWinos\cite{ATLAS:2021yqv} provided new limits on wino masses
depending on the mass of the higgsinos where $m(higgsinos)\sim \mu$.
For instance, for $\mu\sim 250$ GeV, then the range
$m(wino)\sim 0.6-1$ TeV is excluded while for $\mu\sim 150$ GeV then
$m(wino)\sim 0.5-1.02$ TeV is excluded.
For lighter winos with $m\alt 0.5-0.6$ TeV,
then the wino decay products are too soft and the searches become less
efficient.
The LHC limits on $m(wino)$ correspond (in the case of $\mu\sim 250$ GeV)
to limits on $m_{3/2}: 200-360$ TeV.
Meanwhile, the naturalness limit $\Delta_{EW}\alt 30$ occurs for
$m_{3/2}\alt 260$ TeV-- within the LHC-excluded range from
boosted hadronically-decaying EWino searches. Thus, at present there exists
a {\it gap} in nAMSB parameter space $m_{3/2}: 90-200$ TeV where
$m_h\sim 125$ GeV with $\Delta_{EW}\alt 30$ and which is allowed by
present LHC search constraints.

The goal of the present paper is to ascertain if it is possible
for LHC upgrades such as high luminosity LHC ($pp$ collisions
at $\sqrt{s}=14$ TeV with 3000 fb$^{-1}$ integrated luminosity) to
{\it close the gap}, thus either discovering SUSY in its nAMSB
incarnation or else ruling the model out. In pursuit of this goal,
in Sec. \ref{sec:namsb} we first display the nAMSB parameter space in
various parameter planes so as to delineate the remaining target
model parameter space.
Unlike natural SUSY models like the $i$-extra parameter non-universal
Higgs model (NUHMi\cite{Baer:2012cf}) and natural generalized
mirage mediation (nGMM\cite{Baer:2016hfa}),
it appears nAMSB is susceptible to complete
exploration within its natural regions due to the presence of relatively
light unstable winos which lead to several distinct collider search
channels.
In Sec. \ref{sec:BM}, we propose new nAMSB model lines for which we can
explore LHC collider signals and SM backgrounds.
In Sec. \ref{sec:lhc}, we examine several LHC search channels which may help
ATLAS/CMS close the gap.
These include
\bi
\item Soft opposite-sign  dileptons from higgsino pair production,
\item Same sign diboson $W^\pm W^\pm$ signature\cite{Baer:2013yha}
  arising from wino pair production,
\item Hard isolated trileptons from wino pair production and
\item top-squark pair production.
\ei
These channels may be used to fully explore the remaining gap in nAMSB model parameter space
at HL-LHC.
Our summary and conclusions follow in Sec. \ref{sec:conclude}.

\section{Overview of nAMSB parameter space}
\label{sec:namsb}

\begin{figure}[tbp!]
\begin{center}
  \includegraphics[height=0.3\textheight]{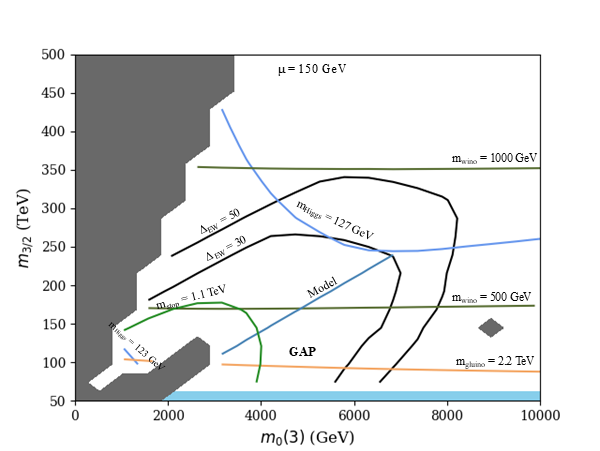}\\
      \includegraphics[height=0.3\textheight]{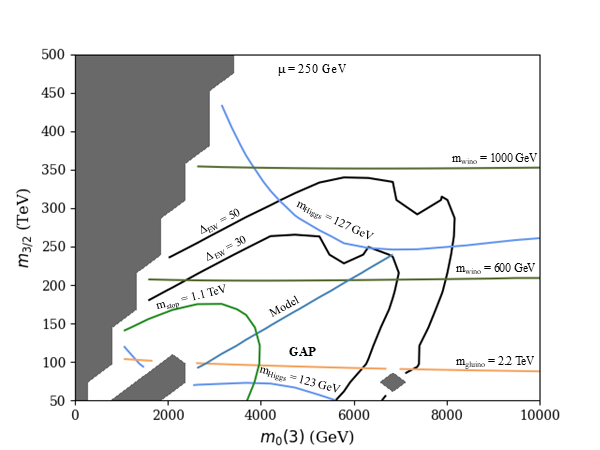}
    \caption{Plot of $m_0(3)$ vs. $m_{3/2}$ parameter space in the
      nAMSB model for $m_0(1,2)=2m_0(3)$, $A_0=1.2 m_0(3)$ and $\tan\beta =10$,
      with $m_A=2$ TeV and {\it a}) $\mu =150$ GeV and {\it b}) $\mu =250$ GeV.
\label{fig:m03m32}}
\end{center}
\end{figure}

In many past works, the mAMSB model parameter space has been displayed in
the $m_0$ vs. $m_{3/2}$ plane\cite{Feng:1999hg,Baer:2000bs,Barr:2002ex}
for particular values of $\tan\beta$.
The same can be performed for nAMSB except now the plane is
sensitive to additional parameter possibilities.
In Fig. \ref{fig:m03m32}, we present nAMSB parameter space in the
$m_0(3)$ vs. $m_{3/2}$ plane where now we take the first/second generation
bulk masses $m_0(1,2)\ne m_0(3)$, with $m_0(1,2)=2m_0(3)$ and bulk trilinear $A_0=1.2m_0(3)$. 
The motivation here comes in part from expectations from SUGRA where non-universal
bulk soft terms are expected, and the string landscape which
would preferentially select large soft terms so long as they do not contribute
too much to the determination of the weak scale (no large weak scale
finetuning). We take each generation of matter scalars as degenerate
since all elements of each generation fill out the 16-d spinor rep of
$SO(10)$. The large first/second generation scalar masses which are expected
actually can provide a landscape mixed decoupling/quasi-degeneracy solution
to the SUSY flavor and CP problems\cite{Baer:2019zfl}.

Fig. \ref{fig:m03m32} is portrayed for $\tan\beta =10$ and $m_A=2$ TeV,
with {\it a}) $\mu =150$ GeV and {\it b}) $\mu =250$ GeV.
We use Isajet 7.91 model 13 to compute spectra for the nAMSB
model\cite{Paige:2003mg}.
The left gray region is where sleptons become the LSP.
Here, we assume $R$-parity conservation which is less ad-hoc
than it used to be in that both $R$-parity and global $U(1)_{PQ}$ can arise
from more fundamental discrete $R$-symmetries which provide
simultaneously solutions both  to the SUSY $\mu$ problem and also the
axion quality problem\cite{Baer:2018avn,Bhattiprolu:2021rrj}.
Thus, the left gray region is excluded. The lower gray region around
$m_0(3)\sim 2000$ GeV is also excluded, this time because stop soft
terms run tachyonic leading to charge and/or color breaking (CCB) minima
in the scalar potential.

In Fig. \ref{fig:m03m32}, we also show contours of $m_h=123$ and 127 GeV.
Thus, the region between the $m_h:123-127$ GeV contours is in accord with
LHC $m_h$ measurements (due to the large bulk $A-0$ terms).
We also show black contours of electroweak naturalness
$\Delta_{EW} =30$ and 50.
The regions below these contours are considered
natural since all independent contributions to the weak scale are comparable to
the weak scale (no finetuning needed for the LHP). The finetuning increases
with $m_{3/2}$ since large $m_{3/2}$ yields a large value of $m_{\tg}$
and $m_{\tst_{1,2}}$ which in turn increases the $\Sigma_u^u(\tst_{1,2})$
contribution to the weak scale.
Also, the region with too large $m_0(3)$ is excluded since the large
stop sector soft terms lead to too large $\Sigma_u^u(\tst_{1,2})$ values.

The region of Fig. \ref{fig:m03m32} below the orange contour
labeled $m_{\tg}=2.2$ TeV (for $m_{3/2}\alt 90-100$ TeV)
is excluded by ATLAS/CMS limits on gluino pair production
searches\cite{Canepa:2019hph,ATLAS:2024lda}.
The region between the two green contours labeled $m_{\tchi_2^+}=500,\ 600$ and
1000 GeV is excluded by the ATLAS search for EWinos which decay to $W$ or $Z$
which in turn decay hadronically\cite{ATLAS:2021yqv}.
This region includes the $\Delta_{EW}=30$ and 50 contours and so,
along with these contours, defines an upper limit to nAMSB parameter space (p-space).
The remaining lower-left gap region
is still allowed by naturalness, by the measured value of $m_h$,
by direct/indirect mixed axion/higgsino-like WIMP dark matter
searches\cite{Baer:2016ucr} and by LHC sparticle search limits.
This remaining gap provides a target for future
LHC upgrade searches for the incarnation of natural SUSY via the nAMSB model. 

\begin{figure}[tbp]
\begin{center}
  \includegraphics[height=0.3\textheight]{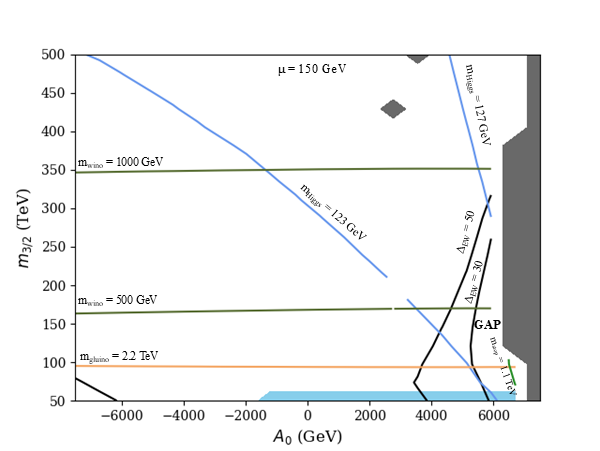}\\
      \includegraphics[height=0.3\textheight]{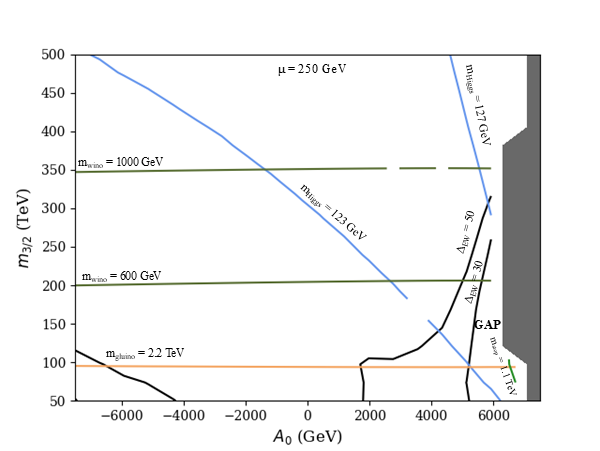}
    \caption{Plot of $A_0$ vs. $m_{3/2}$ parameter space in the
      nAMSB model for $m_0(3)=5$ TeV with $m_0(1,2)=2m_0(3)$
      and $\tan\beta =10$, with $m_A=2$ TeV and {\it a}) $\mu =150$ GeV and
      {\it b}) $\mu =250$ GeV.
\label{fig:A0m32}}
\end{center}
\end{figure}

In Fig. \ref{fig:A0m32} we display the nAMSB parameter space in the
$A_0$ vs. $m_{3/2}$ plane for fixed $m_0(3)=5$ TeV and $m_0(1,2)=10$ TeV,
with $\tan\beta =10$ and $m_A=2$ TeV  with {\it a}) $\mu =150$ GeV
and {\it b}) $\mu =250$ GeV (as before).
Here, we find a right-side gray region where large values of $A_0$ cause
$m_{U_3}^2$ to run tachyonic. The bulk of this p-space plane is unnatural
with $\Delta_{EW}\agt 50$ except for the region to the right of the
naturalness contours. Also, we see that $m_h\sim 125$ GeV mainly on the
right-hand-side unless $m_{3/2}$ is very large, in which case the spectra are
unnatural.
Here it is plain to see the advantage that bulk trilinear terms
provide: for $A_0\sim 0$, then the model is unnatural and yields $m_h$
too light unless $m_{3/2}\agt 300$ TeV.
We also show in the plot the orange contours of $m_{\tg}=2.2$ TeV
(below is excluded)
and $m_{\tchi_2^+}=500,\ 600$ and 1000 GeV (between which is excluded).
The remaining non-excluded target p-space lies on the extreme right
where $A_0\sim 5-7$ TeV.

\section{An $nAMSB$ model line}
\label{sec:BM}

Our goal is to explore the LHC-allowed nAMSB parameter space
with $\Delta_{EW}\alt 30$ to see if it is completely testable at HL-LHC.
To this end, we construct a nAMSB model line that cuts through the bulk of
allowed p-space but which varies with $m_{3/2}$ from the lower
limit formed by LHC gluino pair searches to the upper limit which is formed
by naturalness combined with LHC boosted hadronically-decaying EWino pair
searches. Our constructed model line is shown in Fig. \ref{fig:m03m32}
as the blue diagonal line labeled {\it Model}.
It is parametrized by
\be
m_{3/2}=35 m_0(3)\ \ \ ({\rm nAMSB\ model\ line})
\ee
with $m_0(1,2)=2 m_0(3)$, $A_0=1.2 m_0(3)$ and $\tan\beta =10$ with
$m_A=2$ TeV and $\mu =150$ GeV and 250 GeV.

\begin{figure}[tbp]
\begin{center}
    \includegraphics[height=0.3\textheight]{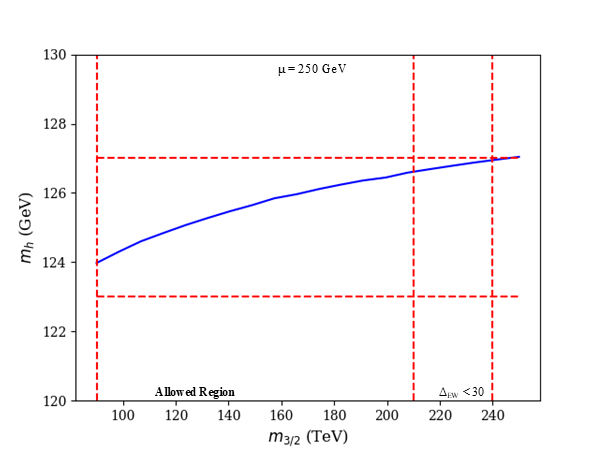}
    \caption{Plot of $m_h$ vs. $m_{3/2}$ along the
      nAMSB model-line for $m_{3/2}=35 m_0(3)$ with $m_0(1,2)=2m_0(3)$,
      $A_0=1.2 m_0(3)$ and $\tan\beta =10$, with $\mu =250$ GeV and $m_A=2$ TeV.
\label{fig:mh}}
\end{center}
\end{figure}

In Fig. \ref{fig:mh}, we plot the value of
$m_h$ along the model line with $\mu =250$ GeV, which extends from
$m_{3/2}=90$ TeV (left-most vertical red-dashed line) to $m_{3/2}\sim 210$ TeV
(where LHC gaugino limits set in, denoted by the dashed red line at
$m_{3/2}\sim 210$ TeV).
From the plot, we see that $m_h$ lies between 123-127 GeV,
and so is consistent with the measured value of $m_h$ up to a theory error
of $\pm 2$ GeV.

\begin{figure}[tbp]
\begin{center}
    \includegraphics[height=0.3\textheight]{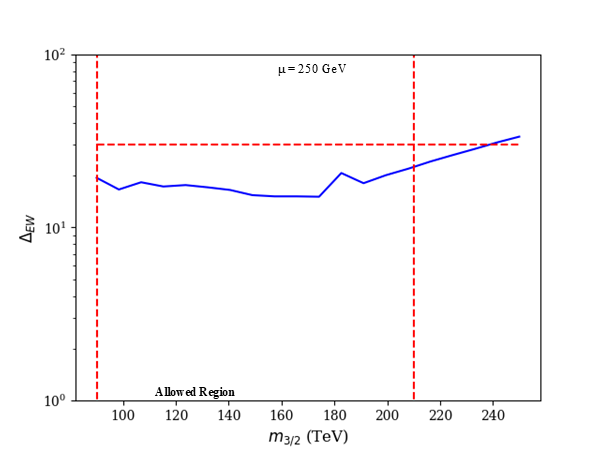}
    \caption{Plot of $\Delta_{EW}$ vs. $m_{3/2}$ along the nAMSB model-line
      for $m_{3/2}=35 m_0(3)$, $m_0(1,2)=2m_0(3)$, $A_0=1.2 m_0(3)$ and
      $\tan\beta =10$, with $\mu =250$ GeV and $m_A=2$ TeV.
\label{fig:dew}}
\end{center}
\end{figure}

In Fig. \ref{fig:dew}, we plot the value of $\Delta_{EW}$ along the
$\mu =250$ GeV model line.
In this case, we find that $\Delta_{EW}$ is typically between values of
15-30 along the line until it climbs beyond 30 (the red-dashed line) for
$m_{3/2}\agt 240$ TeV. 

\begin{figure}[tbp]
\begin{center}
    \includegraphics[height=0.4\textheight]{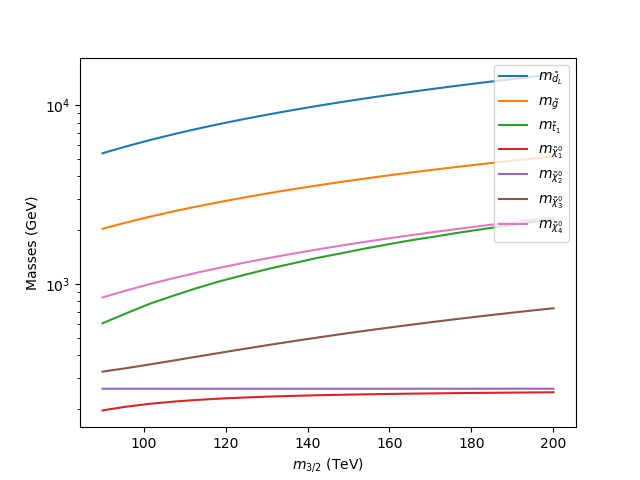}
    \caption{Plot of various sparticle masses vs. $m_{3/2}$
      along the nAMSB model-line for $m_{3/2}=35 m_0(3)$
with $m_0(1,2)=2m_0(3)$, $A_0=1.2 m_0(3)$ and $\tan\beta =10$,
      with $\mu =250$ GeV and $m_A=2$ TeV.
\label{fig:masses}}
\end{center}
\end{figure}

In Fig. \ref{fig:masses}, we plot various sparticle masses along the nAMSB
model line. At the very bottom, the red and purple (nearly overlapping) curves 
denote the higgsino-like neutralinos $\tchi_{1,2}^0$ and their masses are
fixed to be very nearly $\sim \mu =250$ GeV. The next-higher brown curve
denotes the neutral wino $m_{\tchi_3}^0\sim M_2$ and these masses range between
275-600 GeV in the allowed region. Such light winos provide an inviting target
for LHC searches since even higher values are already excluded by
the LHC boosted EWino search results.
The green curve shows the lighter top-squark mass $m_{\tst_1}$ which extends
over the range $m_{\tst_1}: 0.6-2$ TeV.
(The lower part of this range is already excluded by LHC top-squark searches
within simplified models\cite{Canepa:2019hph,ATLAS:2024lda}.)
The bino mass $m_{\tchi_4^0}$ lies just above $m_{\tst_1}$ for this model line.
Meanwhile, the gluino mass varies from $m_{\tg}:2.2-4.5$ TeV while
first/second generation sfermions lie in the 5-15 TeV range.
For our HL-LHC search strategy, we will focus on the relatively light
higgsinos and winos along this model line.

\section{nAMSB discovery channels for HL-LHC}
\label{sec:lhc}

We next examine the reach of HL-LHC ($\sqrt{s}=14$ TeV with 3000 fb$^{-1}$)
for wino pair production in the nAMSB model.
To proceed, we generate a SUSY Les Houches Accord (SLHA)
file\cite{Skands:2003cj} for each of our nAMSB SUSY model line points
using Isajet 7.91\cite{Paige:2003mg} and feed this into
Madgraph\cite{Alwall:2011uj}/Pythia8\cite{Sjostrand:2006za}
which is used for both signal and background (BG) processes.
The $pp\to\tchi_2^\pm\tchi_3^0 X$ cross section is normalized to the
Prospino NLO\cite{Beenakker:1996ed}.
For the $2\to 3$ BG processes, we also use Madgraph\cite{Alwall:2011uj}
coupled to Pythia.
We adopt the toy detector simulation Delphes\cite{deFavereau:2013fsa}
using the default ATLAS card.

In Delphes, we identify the following entities.
\bi
\item Jets are reconstructed using the anti-$k_T$ algorithm with
  $p_T(j)>20$ GeV within a cone of $\Delta R=0.4$. Such clusters are
  labelled as jets when $p_T(j)>40$ GeV with pseudorapidity $|\eta (j)|<3.0$.
\item Leptons with $p_T>10$ GeV are reconstructed in a cone of $\Delta R=0.2$.
  The maximum hadronic energy allowed in the cone is 10\% of the leptonic
  $p_T$. Thus, leptons are identified when $p_T(\ell )>10$ GeV and
  $|\eta (\ell )|<2.5$.  
\ei

\subsection{Same-sign dileptons from wino pair production}
\label{ssec:SSdB}

\begin{figure}[tbp]
\begin{center}
    \includegraphics[height=0.3\textheight]{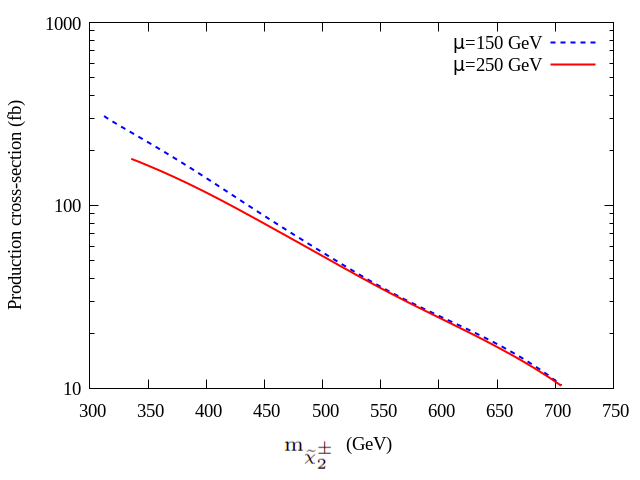}
    \caption{The production cross section  
        from wino pair production $\widetilde{\chi}^{\pm}_2\widetilde{\chi}^0_3$
        along the nAMSB model line with $\mu =150$ and $250$ GeV.
      \label{fig:cs}}
\end{center}
\end{figure}

Here we examine wino pair production $pp\to \tchi_2^\pm\tchi_3^0 X$
along our nAMSB model line at LHC14
where the relevant decay modes are $\tchi_2^\pm\to\tchi_{1,2}^0 W^\pm$ and
$\tchi_3^0\to\tchi_1^\pm W^\mp$.
In this configuration, we expect a final state containing same-sign
dibosons $W^\pm W^\pm +\eslt$ about 50\% of the time leading to a
(relatively) jet-free same-sign dilepton $+\eslt$ signature for which
SM backgrounds  are typically quite low.
The production cross-section for this channel at $\sqrt{s}=14$ TeV LHC
for two values of $\mu=150, 250$ GeV are shown in Fig.~\ref{fig:cs}.
This signal channel- proposed in Ref. \cite{Baer:2013yha} and expanded upon
in Ref. \cite{Baer:2017gzf}-- is endemic to SUSY models with light higgsinos
although so far there are no dedicated experimental analyses.
Here, we will adopt the signal cuts of Ref. \cite{Baer:2017gzf}:
The original cuts from Ref. \cite{Baer:2013yha} are labelled as cut set
{\bf C1} which require:
\bi
\item the presence of exactly two isolated same-sign leptons with
  $p_T (\ell_1)>20$ GeV and $p_T(\ell_2)>10$ GeV, where $\ell_1$ ($\ell_2$)
  is the higher (lower) $p_T$ lepton,
\item veto events with identified $b$-jets,
\item $\eslt >200$ GeV and
\item $min (m_T(\ell_1,\eslt ),m_T ( \ell_2 ,\eslt))>175$ GeV.
\ei
After inspecting distributions for signal/background events,
Ref. \cite{Baer:2017gzf} augmented these with {\bf C2} cuts:
\bi
\item $n(jets)\le 1$ and
\item $\eslt >250$ GeV.
\ei

The backgrounds coming from $t\bar{t}$, $WZ$, $t\bar{t}W$, $t\bar{t}Z$,
$t\bar{t}t\bar{t}$, $WWW$, $WWjj$ $WWZ$, $WZZ$ and $ZZ$ were computed and tabulated in Table II
of Ref. \cite{Baer:2017gzf}. The total background after cuts {\bf C1}
and {\bf C2} (with all BG total cross sections scaled to their NLO results)
was $\sim 8.62$ ab with the largest coming from $WWW$ and $WWjj$
production at 2.21 ab as seen in Table~\ref{tab:ssdlbkg}.  
\begin{table}
 \scriptsize
    \centering
    \begin{tabular}{|c|c|}
       \hline 
       SM Bkg  & Cross-section (after cuts) (ab) \\
       \hline 
    $t\bar{t}$ & 0.0 \\
    $t\bar{t}t\bar{t}$ & 0.0125 \\
    $t\bar{t}W$ & 0.52 \\
    $t\bar{t}Z$ & 0.88 \\
    $WWjj$ & 2.21 \\
    $WWW$ & 2.21 \\
    $WWZ$ & 1.358 \\
    $WZZ$ & 1.426 \\
    $ZZ$ & 0.0 \\
    $WZ$ & 0.0 \\	    
    \hline 
    Total SM background & 8.62 \\
    \hline 
    \end{tabular}
    \caption{The SM background cross-sections after cut \textbf{C2} for the
      same-sign dilepton + $\eslt$ signal at $\sqrt{s}=14$ TeV.}
    \label{tab:ssdlbkg}
\end{table}

\begin{figure}[tbp]
\begin{center}
    \includegraphics[height=0.3\textheight]{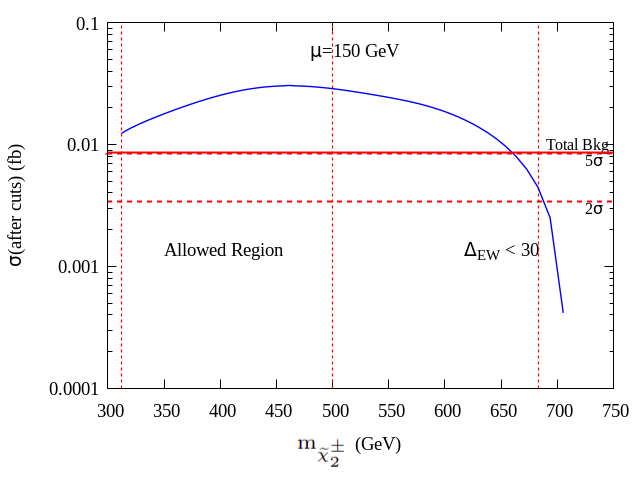}
    \includegraphics[height=0.3\textheight]{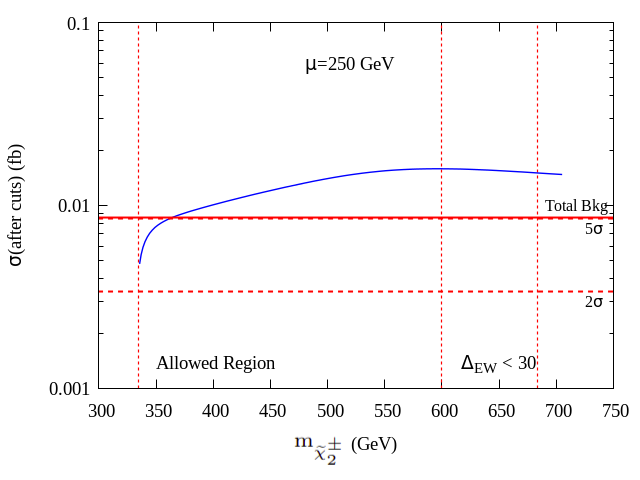}
    \caption{Cross section after cuts for the SSdB signature
      from wino pair production along the nAMSB model line
      for {\it a}) $\mu =150$ GeV and {\it b}) $\mu =250$ GeV.
      We also show the remaining total background rate.
      \label{fig:ssdl}}
\end{center}
\end{figure}

The signal cross section after cuts {\bf C2} is shown vs. $m_{\tchi_2^+}$
along the nAMSB model line, for {\it a}) $\mu =150$ and {\it b}) $250$ GeV
in Fig. \ref{fig:ssdl}.
In Fig. \ref{fig:ssdl}{\it a}), we denote the total SM background by the
solid red horizontal line at 8.6 ab while the HL-LHC 3000 fb$^{-1}$
$5\sigma$ line is horizontal red-dotted as is the 95\%CL line.
The left-most vertical line denotes the
parameter space limit from LHC gluino pair searches where
$m_{\tg}> 2.2$ TeV is required. The red-dotted vertical line
at $m_{\tchi_2^\pm}=500,\ 600$ GeV denotes the upper limit on p-space
from the ATLAS boosted hadronic jet search
from wino pair production\cite{ATLAS:2021yqv}.
Beyond the right-most vertical red-dotted line is where
the naturalness measure $\Delta_{EW}$ exceeds 30. From frame {\it a}), we
see that for $\mu =150$ GeV almost the entire presently-allowed p-space gap
should be accessible to HL-LHC via the SSdB signal channel,
well above the $5\sigma$ discovery limit (the exception being a tiny slice with low $m_{3/2}$
where decay products are rather soft and don't exceed our rather hard cuts; this small
slice may be filled by updated gluino pair search results).

In Fig. \ref{fig:ssdl}{\it b}), we plot the SSdB signal rates for the same model
line parameters except that $\mu= 250$ GeV.
In the $\mu =250$ GeV case,
we see that the bulk of presently-allowed p-space can be seen by
HL-LHC at the $5\sigma$ level, except for the region with
$m_{\tchi_2^\pm}\alt 360$ GeV. For this lower range of wino mass values,
the wino decays into real $W$s are still open, but the wino
decay products are much softer, and a softer $\eslt$ cut may be necessary
to pick out the lower mass wino pair events. In any case, the signal
is only slightly below our projected $5\sigma$ level, and well above the
95\%CL exclusion level.

\subsection{Hard isolated trileptons from wino pair production}
\label{ssec:hard3l}

In this Subsection, we focus on wino pair production $pp\to\tchi_2^\pm\tchi_3^0$
followed by decays $\tchi_2^\pm\to W^\pm\tchi_{1,2}^0$ and
$\tchi_3^0\to Z^0\tchi_{1,2}^0$. When both $W$ and $Z$ decay leptonically, then
we expect the clean trilepton $+\eslt$ final state topology\cite{Baer:1985at,Baer:1994nr}.

We evaluate the signal and SM background using Madgraph/Pythia/Delphes as
before. After examining signal and background distributions,
we the implement the following cuts
\bi
\item The final state contains three identified isolated leptons with
  $p_T(\ell_1 )>20$ GeV, $p_T(\ell_2)>15$ GeV and $p_T(\ell_3)>10$ GeV,
  where $\ell_1$ is the highest $p_T$ lepton, and so forth.
  \item  $b$-jet veto: $n_{b-jet}=0$ and $\gamma$ veto: $n_{\gamma}=0$.
\item $\eslt >100$ GeV.
\item Effective mass $m_{eff}>200$ GeV (where $m_{eff}$ is the scalar sum
  of the $p_T$s of the three leptons plus the $\eslt$).
\item Transverse mass $m_T(\ell_i ,\eslt)>120$ GeV for each isolated
  lepton $i=1-3$.
\item Jet veto: $n_j=0$.

\ei
The various SM cross sections
(normalized to their NLO values, and where available, up to NNLO+NNLL)
after these cuts are listed in Table \ref{tab:3lBG}.
The dominant backgrounds are summarized as below:
\begin{itemize}
    \item $t\bar{t}$ - NNLO + NNLL \cite{Czakon:2011xx}
    \item $WZ, ZZ,$ - NLO \cite{Campbell:2011bn}
    \item $t\bar{t}Z, t\bar{t}W$- NLO (QCD + EW)\cite{LHCHiggsCrossSectionWorkingGroup:2016ypw}
    \item $WWZ, WWW, ZZZ$ - NLO \cite{Alwall:2014hca}
    \item $tZj$ - NLO QCD+EW \cite{Pagani:2020mov}
    \item $tW$ - NNLO \cite{Kidonakis:2015nna}
\end{itemize}
\begin{table}[ht]
\begin{center}
\scriptsize
\begin{tabular}{|c|c|}
      \hline
      SM background & Cross-section after cuts (ab)\\
      \hline
         $t\bar{t}$  &  2.92\\
       $WZ$  & 0.712\\
        $ZZ$&0.45\\
         $ttZ$&2.25\\
          $ttW$&0.32\\
          $WWZ$  & 5.12\\
          $WWW$  & 0.035\\
          $ZZZ$ & 0.0047\\
           $tZj$&0.158\\
           $tW$ & 0\\
            \hline 
         Total Background & 11.97 \\ 
        \hline
    \end{tabular}
    \end{center}
\caption{Cross section after cuts in {\it ab} for the dominant SM backgrounds
to the hard $3\ell$ signal at LHC14.}
\end{table}  
\label{tab:3lBG}

\begin{figure}[tbp]
\begin{center}
    \includegraphics[height=0.3\textheight]{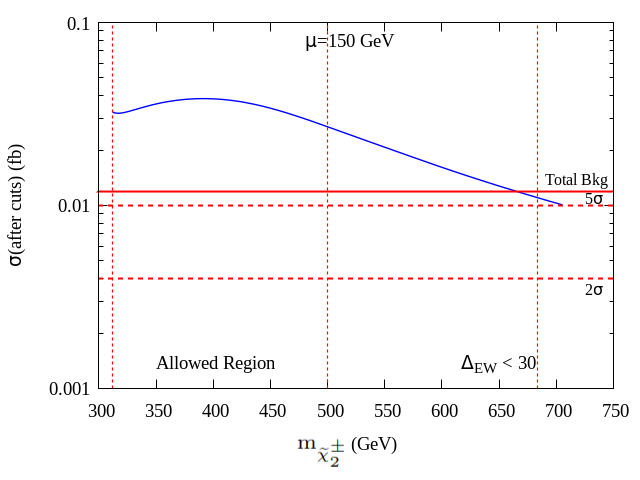}
    \includegraphics[height=0.3\textheight]{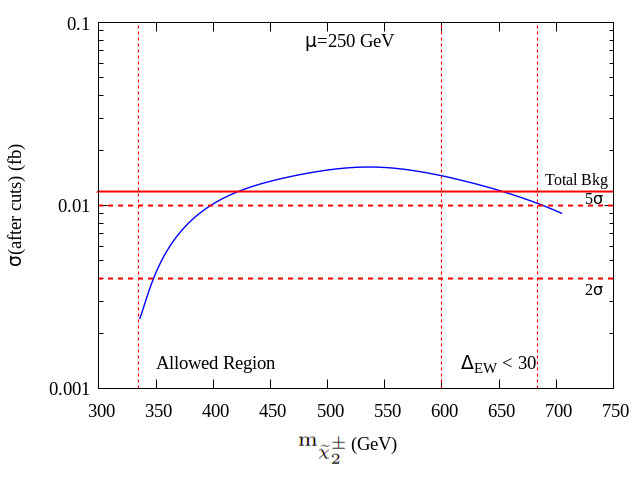}
    \caption{Cross section after cuts for the hard $3\ell$ signature
      from wino pair production along the nAMSB model line
      with {\it a}) $\mu =150$ GeV and {\it b}) $\mu =250$ GeV.
      We also show the remaining total background rate.
      The central vertical dotted line corresponds to the LHC-allowed
      upper limit on p-space.
      \label{fig:3l}}
\end{center}
\end{figure}

In Fig. \ref{fig:3l}{\it a}), we show a plot of the cross section after cuts
along the nAMSB3 model line for $\mu=150$ as a
function of $m_{\tchi_2^\pm}$, along with the total remaining SM background.
We also show the $5\sigma$ and $95\%CL$ reach of LHC14 for 3000 fb$^{-1}$
of integrated luminosity. In this case, the wino pair production
clean $3\ell$ signal exceeds the $5\sigma$ level for
$m_{\tchi_2^\pm}\sim 300-620$ GeV, which is the entire allowed gap.
A signal in this channel would provide strong confirmation
of any signal already present in the SSdB channel.

In Fig. \ref{fig:3l}{\it b}), we show the hard $3\ell$ signal along the
nAMSB model line but with $\mu =250$ GeV. In this case, the signal
exceeds the $5\sigma$ level for the more limited range of
$m_{\tchi_2^\pm}\sim 450-600$ GeV. Thus, the lower portion of the gap is more
difficult to access due to the softer wino decay products, since the wino
decay to real $W$ bosons is just barely open. Nonetheless, the signal
exceeds the 95\%CL range except for the very lowest values of
$m_{\tchi_2^\pm}\alt 350$ GeV,

\subsection{Soft OS dileptons from higgsino pair production}
\label{ssec:osdjmet}

Next we turn to nAMSB signals from higgsino pair production. The most
lucrative reaction in this case comes from\cite{Baer:2011ec}
\be
pp\to\tchi_1^0\tchi_2^0\ \ \ \ {\rm with}\ \ \ \tchi_2^0\to\tchi_1^0\ell\bar{\ell}.
\ee
The visible decay products are very soft since the mass gap
$m_{\tchi_2^0}-m_{\tchi_1^0}$ tends to be in the 5-50 GeV range, and most of the
$\tchi_2^0$ decay energy release goes into making the $\tchi_1^0$ rest mass.
The final state lepton energy can be boosted to higher values, and events can be triggered upon, by requiring a hard ISR jet emission\cite{Han:2014kaa,Baer:2014kya}. 
This is then the soft OS dilepton plus jet plus MET signature (OSDLJMET).
Five selection cuts plus a ditau reconstruction cut requiring
$m_{\tau\tau}^2<0$ to reject backgrounds such as $Zj$ production
(followed by $Z\to\tau\bar{\tau}$) were required in Ref. \cite{Baer:2014kya}.
In Ref. \cite{Baer:2021srt}, more efficient angular cuts were developed which
reduce the ditau backgrounds to tiny levels. The original cut set {\bf C1}
augmented by new angular cuts to reject $Zj\to \tau\bar{\tau}j$ plus further
cuts to reject $t\bar{t}$ and $WW$ backgrounds were labelled set {\bf C3}
in Ref. \cite{Baer:2021srt,Baer:2022qrw}. After cut-set {\bf C3}, signal and
SM BG were plotted against opposite-sign dilepton invariant mass
$m(\ell\bar{\ell})$ where signal should accrue for
$m(\ell\bar{\ell})<m_{\tchi_2^0}-m_{\tchi_1^0}$ whilst for higher invariant masses
data should be in accord with SM expectations.

\begin{figure}[tbp]
\begin{center}
  \includegraphics[height=0.3\textheight]{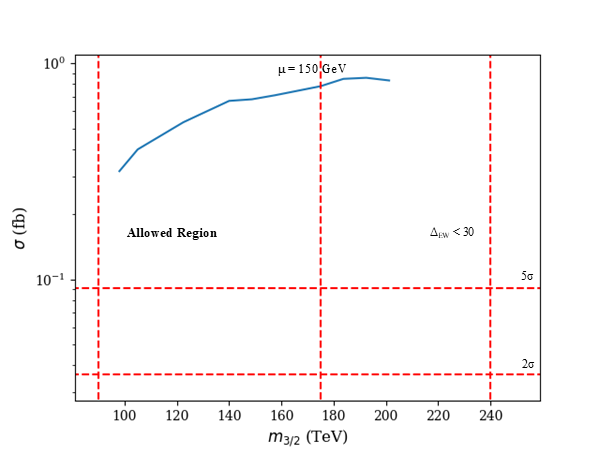}\\
      \includegraphics[height=0.3\textheight]{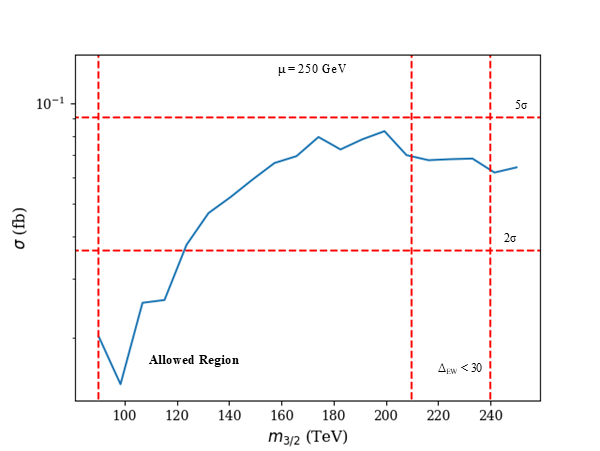}
    \caption{Cross section after cuts for the soft opposite-sign
      dilepton plus jet plus $\eslt$ events
      from higgsino pair production along the nAMSB model line
      with {\it a}) $\mu =150$ GeV and {\it b})  $\mu =250$ GeV.
      We also show the $5\sigma$ and $95\%$CL lines for HL-LHC with
      3 ab$^{-1}$ of integrated luminosity.
      \label{fig:osdljmet}}
\end{center}
\end{figure}

In the present work, we evaluate signal along the nAMSB model lines for
$\mu =150$ GeV and $250$ GeV using cuts {\bf C3} along with
the dilepton invariant mass edge cut and compare to SM backgrounds as
evaluated in Ref's \cite{Baer:2021srt,Baer:2022qrw}. Our results are shown
in Fig. \ref{fig:osdljmet}{\it a}) and {\it b}) for $\mu =150$ and 250 GeV,
respectively. From Fig. \ref{fig:osdljmet}{\it a}) for $\mu =150$ GeV,
we see the blue signal curve lies well above the $5\sigma$ BG level for
LHC14 with 3 ab$^{-1}$ over the whole gap region. However, in frame {\it b})
for $\mu =250$ GeV, the higgsino pair production signal is below the
$5\sigma$ level for the entire allowed range of $m_{3/2}$ although it is
above the 95\%CL level for $m_{3/2}\sim 125-210$ TeV.

\begin{figure}[tbp]
\begin{center}
  \includegraphics[height=0.3\textheight]{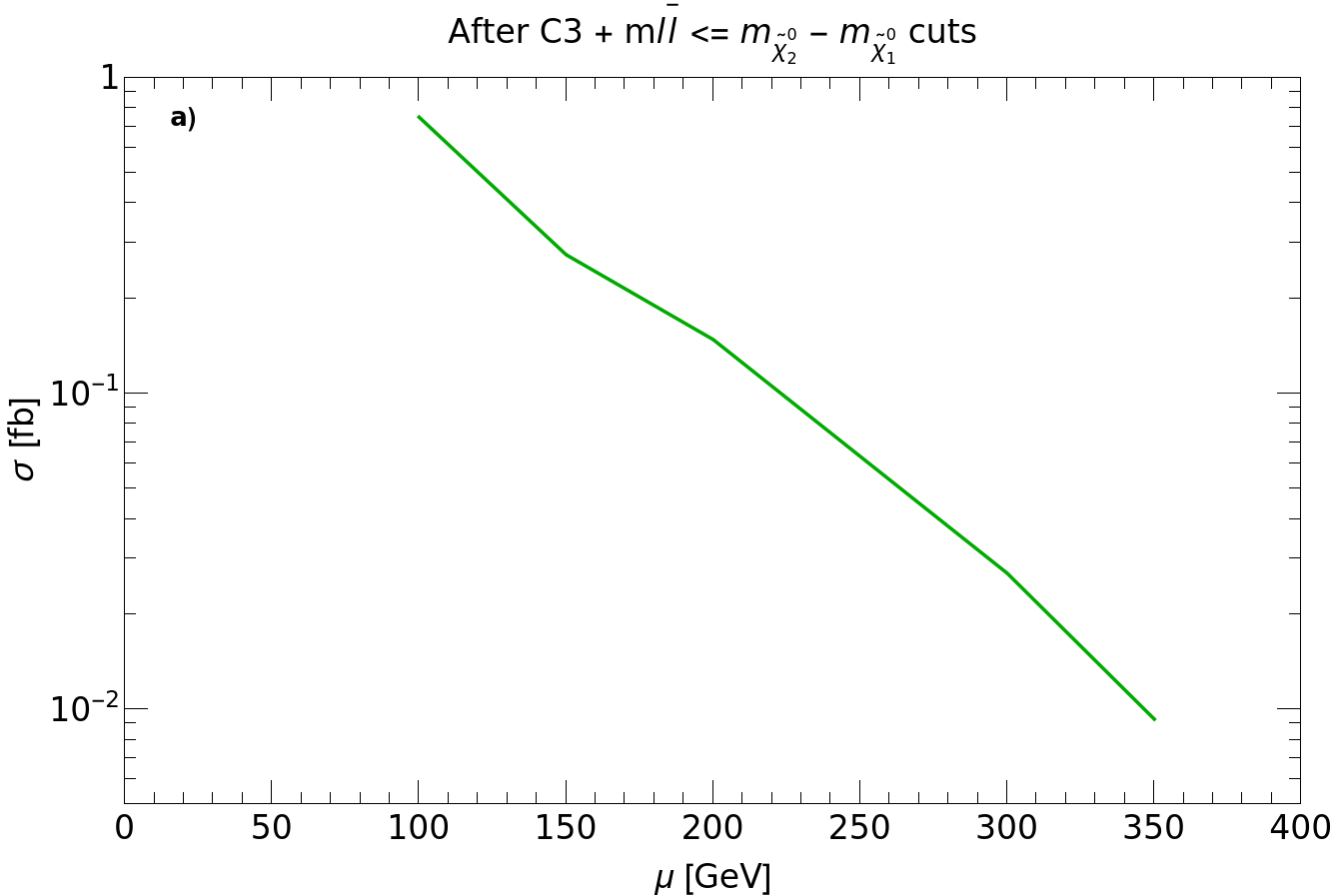}\\
      \includegraphics[height=0.3\textheight]{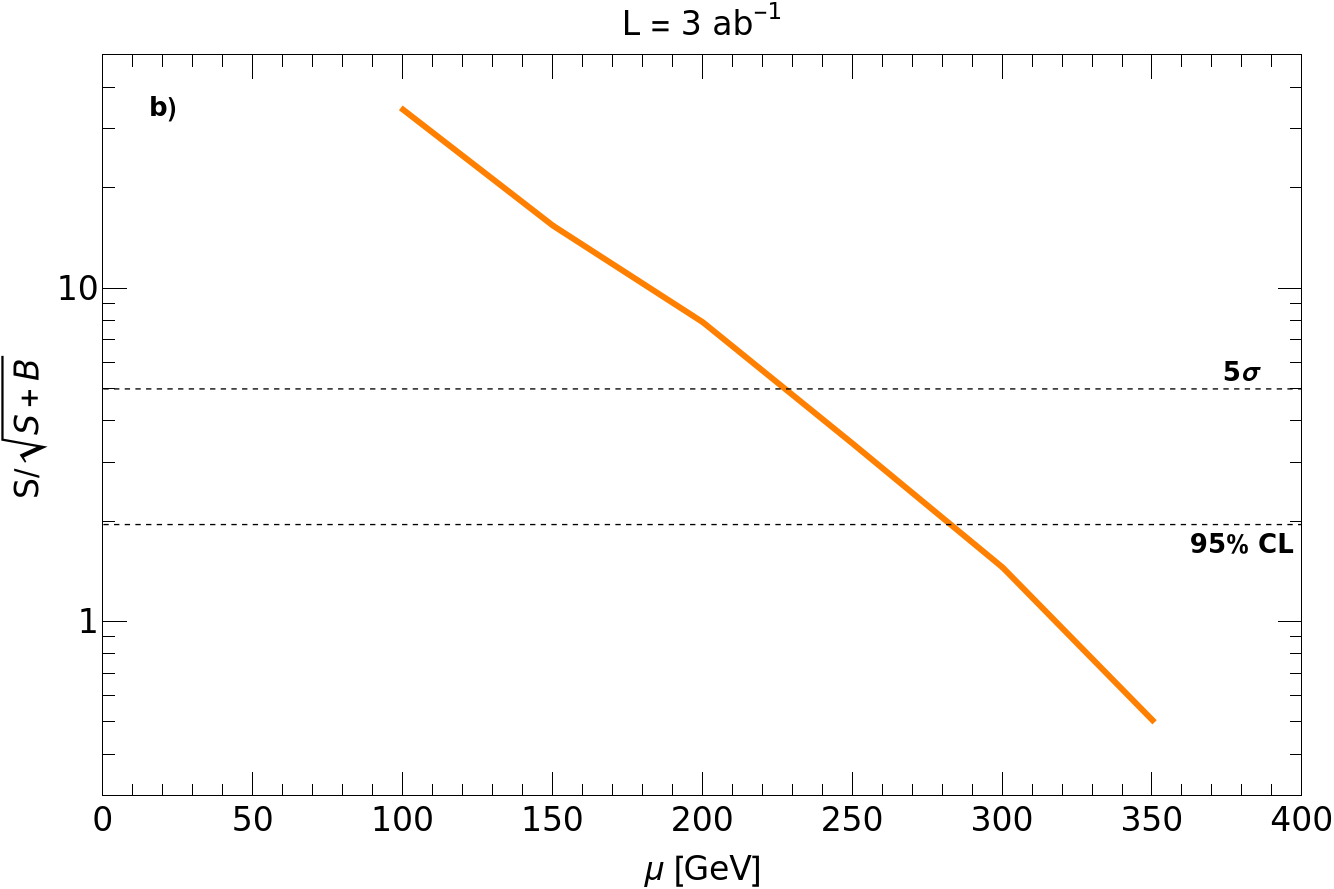}
    \caption{Cross section after cuts for the soft opposite-sign
      dilepton plus jet plus $\eslt$ events
      from higgsino pair production along the nAMSB model line
      vs. $\mu$ but with with $m_{3/2}=150$ TeV, $m_0(3)=m_{3/2}/35$,
      $m_0(1,2)=2m_0(3)$, $A_0=1.2m_0(3)$, $m_A=2$ TeV and $\tan\beta =10$.
      In {\it a}), we show remaining signal cross section after cuts
      {\bf C3} of Ref. \cite{Baer:2021srt} but with
      $m(\ell\bar{\ell})<m_{\tchi_2^0}-m_{\tchi_1^0}$, and in frame {\it b}),
      we show the $n$-$\sigma$ value after cuts
      at LHC14 with 3 ab$^{-1}$ of integrated luminosity.
      \label{fig:sigvsmu}}
\end{center}
\end{figure}

In Fig. \ref{fig:sigvsmu} we show in frame {\it a}) the cross section for the
OSDJMET signal after cuts {\bf C3} of Ref. \cite{Baer:2021srt} and with
$m(\ell\bar{\ell})<m_{\tchi_2^0}-m_{\tchi_1^0}$.
As $\mu$ increases, while $M_2$ is fixed in this plot, then the light higgsinos
become increasingly mixed and the mass gap $m_{\tchi_2^0}-m_{\tchi_1^0}$ also
increases so that the BG increases as well. But for 3 ab$^{-1}$ of
integrated luminosity, the number of signal events remain above 30 even for the
maximal value of $\mu\sim 350$ GeV. In frame {\it b}), we show the
$n$-$\sigma$ level for the OSDJMET signal after cuts for LHC14 with 3 ab$^{-1}$
of integrated luminosity. Here we show that the OSDJMET signal is above the
$5\sigma$ discovery level out to $\mu\sim 225$ GeV and it lies above the
95\%CL exclusion level for $\mu\alt 280$ GeV. For higher $\mu$ values,
a signal in the OSDJMET channel would be difficult to ascertain.

\subsection{Top-squark pair production}
\label{ssec:stop}

For light top-squarks, along the nAMSB model lines we have the
$\tst_1\to t\tchi_{1,2}^0$ and $b\tchi_1^+$ decay modes open, as
is usual in natural models with gaugino mass unification, such as NUHM3.
The difference between nAMSB and NUHM3 is that for nAMSB, the decay
modes $\tst_1\to b\tchi_2^+$ and $\tst_1\to t\tchi_3^0$ may also be open
at the 5-10\% level, leading to some differences from expectations for
NUHM3. The reach of HL-LHC for light top squarks $pp\to\tst_1\tst_1^*X$
was computed in Ref. \cite{Baer:2023uwo} and the $5\sigma$ reach was
found to extend to $m_{\tst_1}\alt 1.7$ TeV and the 95\%CL reach to
$m_{\tst_1}\alt 2$ TeV. We do not expect the results for the reach along
the nAMSB model line to differ much from the NUHM3 model projections.

\section{Summary and conclusions}
\label{sec:conclude}

Anomaly-mediated gaugino masses are expected in a class of models
where some sort of sequestering of tree-level gaugino masses occurs, such
as in models with charged SUSY breaking (no hidden sector singlets\cite{Giudice:1998xp}). 
In contrast, in such models
scalar masses are expected to occur at tree-level with $m_\phi\sim m_{3/2}$
since any remnant hidden sector symmetries do not suppress these
in the K\"ahler potential. While the original mAMSB model now seems excluded
by naturalness bounds combined with requiring $m_h\sim 125$ GeV and
bounds on wino-like WIMP dark matter, an alternative generalized AMSB model
which is characterized by non-universal bulk scalar masses and bulk $A$-terms
is fully allowed. The nAMSB model still has winos as the lightest gauginos,
but now higgsinos are the lightest EWinos, in accord with naturalness.
The presence of rather light winos in this class of models has already led
to strong bounds on the higher range of wino masses from LHC gaugino
searches with hadronically-decaying boosted $W$s and $Z$s, the excluded
range of which encompasses the $\Delta_{EW}\alt 30$ naturalness limit.
This leaves an allowed parameter space gap between lower limits
on $m_{3/2}$ from LHC gluino pair production searches to upper limits
from the boosted hadronically decaying $W,Z$ searches.

We investigated here whether or not this intermediate gap of allowed
p-space can be fully probed at HL-LHC. Our answer is that yes it can,
mainly based on the light winos expected in nAMSB which give rise to
a distinctive SSdB signal (leading to hadronically quiet same-sign dileptons
plus missing transverse energy)
with very tiny SM backgrounds. Usually this signal should exceed the
$5\sigma$ discovery level unless higgsinos are on the heavy side,
whence the visible final state energy is degraded.
Specialized cuts for this regime may boost it back to the $5\sigma$
discovery range.
Along with the SSdB signal, over much of p-space a hard $3\ell$ signal
arising from wino pair production should also be visible. Also, for
not-too-heavy higgsinos, then the soft dilepton plus jet signal from
higgsino pair production should be visible.
Top-squarks with mass $m_{\tst_1}\alt 1.8$ TeV may also be visible to
HL-LHC.
Taken together, it seems
HL-LHC can either discover or at least rule out nAMSB.
This is different from other natural models such as NUHM3 with unified gaugino
masses or generalized mirage mediation with compressed gaugino masses
where it seems sparticles may lie beyond HL-LHC reach while still
fulfilling the naturalness requirment of
$\Delta_{EW}\alt 30$\cite{Baer:2020kwz}.
%
%

\section*{Acknowledgments}

HB gratefully acknowledges support from the Avenir foundation.
VB gratefully acknowledges support from the William F. Vilas Estate.


\bibliography{namsb3}
\bibliographystyle{elsarticle-num}

\end{document}